# Exact Thermodynamics of a Polymer Confined to a Lattice of Finite Size

Edmund A. Di Marzio[*#]
Bio-Poly-Phase, 14205 Parkvale Road, Rockville, MD 20853
Charles M. Guttman[#]

Abstract

We write exact equations for the thermodynamic properties of a linear polymer molecule confined to walk on a lattice of finite size. The dimension of the space in which the lattice resides can be arbitrary. We also calculate polymer density. The boundary can be of arbitrary shape and the attraction of the monomers for the sites can be an arbitrary function of each site. The formalism is even more general in that each monomer can have its own energy of attraction for each lattice site. Multiple occupation of lattice sites is allowed which means that we have not solved the excluded volume problem. For one dimension we recover results obtained previously. The 2-d solution obtained here also solves the problem of an infinite parallelepiped. The method is easily extended by the methods of a previous paper to treat the problem of polymer stars or of branched polymers confined within a finite volume. This exact matrix formalism results in sparse matrices with approximately zM non-zero matrix elements where z is the lattice coordination number and the linear dimension M of the Matrix is equal to the number of lattice sites.

**I Introduction:** Wherever there are polymers there are necessarily polymer interfaces so the problem of polymers at interfaces is ubiquitous. Since polymers are themselves ubiquitous (they appear practically everywhere) the problem of polymers at interfaces is a doubly important problem. A measure of importance is obtained by googling the phrase (polymer OR biopolymer) AND (confined OR interface OR surface). This results in 76,600,000 hits. Obviously it is a impossible task to adequately reference such a vast number of works; we will not even try. In this paper we will treat only the isolated polymer problem. Wide ranging applicability of our treatment will occur only if it can be combined with a self consistent treatment of the many-interfering-polymer-molecules problem, much as was done by Scheutjens and Fleer[1] with their extension of the one dimensional solution of the polymer between two plates problem[2]. A recent review of approaches to treating polymers self-consistently is by Fredrickson[3]. It perhaps appropriate however to provide a short, and by no means exhaustive, list of polymer problems that involve in a more of less critical way knowledge of the polymer interface. See Table 1

It is no wonder then that the polymer interface problem in its many forms has attracted wide attention by theorists and experimentalists alike. An interesting useful recent theoretical work is the paper by Freed et al[4]. It contains references to recent theoretical works as well. Many computer calculations by various authors probably are most effective at obtaining an understanding of the polymer interface and the confined polymer problem. A recent review of computer methods is the paper by Khalatur et al[5].

Never-the-less the existence of any exact treatment, such as the one of this paper, offers added insights that sometimes are a good guide to directions of future research. Perhaps the earliest exact treatment of the polymer at an interface with attraction to the interface was the work of Di Marzio and McCracken[6]. It was valid only for the body centered cubic lattice but it yielded both a true second-order thermodynamic phase transition (in the Eherenfest sense) as well as the notion of a depletion thickness (dearth of monomers at the surface even if one end is covalently tied to the surface)[6]. A subsequent exact solution of an isolated polymer molecule confined between two parallel plates[2] tells us two important things, first the introduction of polymer-surface interaction leads to exactly solved phase transition phenomena for various lattices[7]. One learns much when a phase transition can be solved exactly. Second, it may be possible to self-consistently extend the treatment that we will develop in this paper to polymers competing for the same space since this was done previously and successfully for the parallel plate problem[1]. This would be its greatest use.

In this paper we solve exactly the problem of one linear polymer molecule of N monomer units that can walk on lattices of one, or two, or three dimensions by taking steps to nearest neighbors only. As observed previously the one dimensional solution also solves the polymer between two plates problem[2]. As we will show here the two dimensional solution also solves the infinite parallelepiped problem as well.

We do not solve the excluded volume problem, which means that multiple occupancy of a lattice site by monomers is allowed. On the other hand we allow the strength of interaction of a monomer for a lattice site to be both site specific and monomer specific. This allows for several useful generalizations as described below.

**II Theory:** We will work with a two dimensional square lattice of size 3x5. As we work through the solution it will become obvious that the method generalizes simply to 1) any dimension, 2) different lattices, 3) lattices of any size, and of any degree of openness, 4) arbitrary attractive energies of the monomers for the lattice sites. 5) each monomer can have its own strengths of attraction for the various lattice sites. This arises because each monomer has its own associated matrix.

In Figure 1 we label the 15 lattice sites of our 3x5 square lattice, onto which we allow a polymer to walk, from 1 to 15. Any enumeration scheme is permissible. Now consider the 15x15 matrix of Figure 2. Each column has the statistical weight of the site represented by the column. For example column 7 has the weight $w_7$ but only at those row locations corresponding to its possible nearest neighbors. For site 7 these locations are 2, 6, 8, and 12. Now if we were to multiply a column vector with a 1 in position 7 and a zero in the other 14 positions by the matrix (we use the convention row on left multiplies column on right) then we would obtain a new column vector with $w_7$ at locations 2, 6, 8 and 12. We now multiply this new vector by the matrix again. To envisage what happens it helps to imagine the new column vector to be written as a sum of 4 column vectors, one with zeros everywhere except for a $w_7$ at location 2, one with $w_7$ only at location 6, one with $w_7$ only at location 8, and one with a $w_7$ only at location 12. It is now obvious that we have, by multiplying the matrix twice, properly accounted for taking two steps starting at location 7. If we multiply by the matrix again we will have counted all of the ways one can have taken three steps, and so on.

Multiplying by the matrix a total of N-1 times corresponds to taking N-2 steps on the lattice. Generally the resulting column vector has entries at each position. If want to determine the partition function Q (sum over states) for the polymer starting at lattice site k and ending at lattice site j we simply multiply $W^{N-1}$ from the left by a row vector $P(..0..w_j..0..)^T$ with $w_j$ at location j and zeros elsewhere and from the right by a column vector $P_k$ with 1 at location k and zeros everywhere. Thus

$$Q = (P(..0..w_j..0..))^T W^{N-1} P_k \quad , \quad \text{Starts at k and ends at j.} \tag{1}$$

N-1 steps corresponds to N monomers. This is the reason for $w_j$ in the row vector of Eq. 1. It is the weight of the last monomer of the polymer chain. Notice that Q can also be given by the transpose of the RHS of Eq. 1.

$$Q = (P_k)^T (W^T)^{N-1} P(..0..w_j..0..) \quad , \quad \text{Starts at j and ends at k.} \tag{1A}$$

where $P(..0.. w_j.. 0..)$ is a column vector. The one dimensional treatment given previously worked with the transpose[2]. When serially multiplying the matrices of Eq.1 the weight you carry is the site from which you are stepping, while when serially multiplying the matrices of Eq. 1A one assigns a weight to the site you are stepping onto.

If we want the partition function to represent a stepping initially from location k and ending anywhere on the lattice Then we simply multiply on the left by a row vector $(P(w_j))^T$ which contains $w_j$ at each location j. If we wish to start anywhere and end up anywhere we simply sandwich $W^{N-1}$ with the row vector $(P(w_j)^T)$ on the left and the unit column vector $P_0$ on the right. Then the appropriate formula for the partition function of a N-1 step polymer walking from site k to site j on a lattice of coordination number z is seen to be

$$Q = (P(w_j))^T W^{N-1} P_0 \tag{2}$$

The weight $w_j$ in each case is given by the Boltzmann exponential

$$w_j = \exp(-\varepsilon_j / kT) \tag{3}$$

where $\varepsilon_j$ the attractive energy of the lattice site for the polymer segment.

The connection with thermodynamics is given by the usual formula for the extensive variables Helmholtz free energy F, entropy S and energy U.

$$F = -kT \ln Q = U - TS \quad , \quad S = -\partial F / \partial T \tag{4}$$

It now seems so obvious what to do that it is perhaps useful to just list the generalities.

[1] The linear dimension of the matrix is always equal to M, the number of lattice sites. The number of entries in the matrix is twice the number of bonds connecting the lattice sites since each bond can be traversed in two directions. The matrices are therefore quite sparse since the fraction of non-zero entries in the matrix is of the order of z/M, where z is the coordination number of the lattice. There is a large literature on sparse matrices.

[2] The method works in any dimension. Previously the problem was solved in one dimension[2,7], but the method works in any dimension. The one-dimensional treatments are now significantly generalized by use of $w_{j,k}$. See comment 9 below.

[3] The only restriction on our lattice is that each lattice site be connected to other sites by at least one path. Our lattice can thus be an open structure with some sites just remaining vacant. This can also be achieved by allowing some of the $w_j$ to be zero but it is probably better to just view such sites as empty since the size of the matrix is thereby minimized.

[4] The problem of a confined polymer molecule is solved in principle. The boundary of the confinement can be chosen however we wish (we do not need to have plane surfaces for example) and the boundary sites can be given whatever energies we deem appropriate.

[5] One can also in principle treat the problem of a confined polymer in an external field. For example an electric field can be approximated by making the site energies equal to the electromagnetic potential at that site.

[6] One can treat the problem of different lattices. Thus the work of Rubin[7] who calculated adsorption profiles of a polymer on a surface for various lattices can be generalized to 2 and 3 (and indeed arbitrary) dimensions. Previously we had showed that a second order phase transition in the Ehrenfest sense (discontinuity in the slopes of the S(T) and U(T) curves but not in the functions themselves) occurs for the problem of a polymer between two plates. It would be interesting to see what the effect of increased dimensionality is.

[7] One can treat the case of rectangular parallelepipeds for the simple cubic lattice problem. For example the 3x5 lattice discussed above can be transformed to the problem of a parallelepiped of 3x5 cross-section extending infinitely in both directions by placing the number $2w_j$ along the diagonal of the matrix of figure 2 and using $w_j = \exp(-\varepsilon_j/kT)$ as before. This has the effect of reproducing the 3x5 lattice indefinitely in the orientation perpendicular to the lattice, and it is the new object that the polymer now walks on. The number 2 accounts for the 2 ways we can walk perpendicularly to the plane containing the 3x5 lattice sites. Although in this way we are treating a 3-dimensional problem the linear dimension of the matrix remains the same as before. A slight generalization[8] allows us to treat other lattices.

[8] A quantity of interest is the polymer density as a function of position within the confining walls. For the jth site this is achieved by multiplying each non-zero element of the jth column of the matrix of figure 2 by a marker $\exp(\theta_j)$. The expected number $<v_j>$ of monomers on site j is then given by

$$<v_j> = \partial \ln Q / \partial \theta_j \tag{5}$$

where we have used Eq. 2. If we were using the transpose of Eq. 2 then we would have multiplied the row of the matrix by $\exp(\theta_j)$ as was done in reference 2.

[9] Each monomer unit along the chain can have its own unique affinities for the lattice,

$$w_{j,k} = \exp(-\varepsilon_{j,k}/kT) \tag{6}$$

Here the subscript j labels the column of the matrix and the subscript k labels the location of the monomer unit in the chain starting with 1 and ending with N-1. Instead of $W^{N-1}$ in Eq.1 we would have

$$W^{N-1} \rightarrow \prod W_k \tag{7}$$

where $W_k$ is the W matrix with $w_{j,k}$ replacing $w_j$. The row vector of Eq. 1 also has the replacement of $w_j$ by $w_{j,k}$. Notice that the order of the matrices is now important. This generalization has two uses. First, it can be of use to treat those polymers whose monomers are not all the same such as random copolymers, block copolymers and the various biological polymers whose monomers are invariably diverse. Second, one can now use the method of Guttman et al[9] to treat star molecules and branched molecules. For star molecules the partition function is just the product of the partition functions of each arm provided we start each arm from a common point.
[10] Finally we expect that these results can be used to treat self-consistently the problem of polymers competing for space, just as was done by Scheutjens and Fleer[1] for the effectively one-dimensional case of polymers between two plates or on a one dimensional plane surface.

The above formulation of the confined polymer problem will be of value only if sparse matrices can be handled successfully on the computer. A cube of 30 lattice sites on a side means we are dealing with a large sparse matrix of linear dimension 30x30x30 = 27000. Even the parallelepiped problem would involve a matrix of linear dimension 30x30=900. So it is clear that progress can be made only if we can treat these sparse matrices conveniently. We will address this problem in a future publication which will be submitted to J. Chem. Phys.

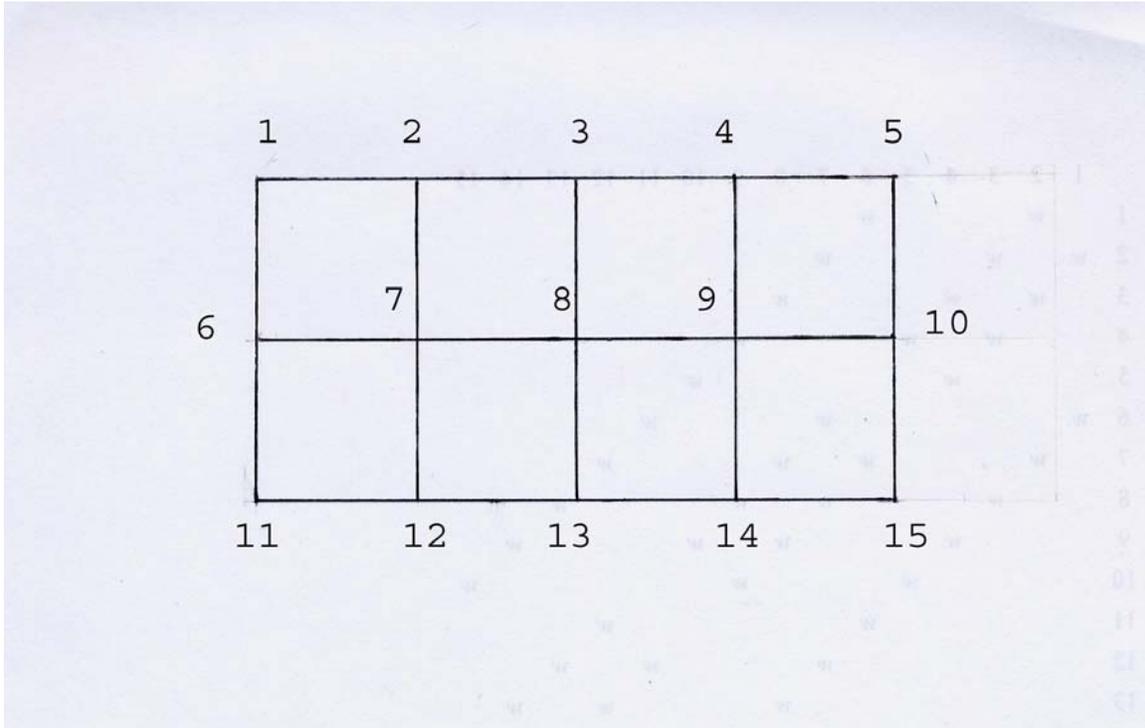

**Figure 1:** For clarity of presentation we work on a square lattice of 3x5 sites. We label the sites as shown but any labeling scheme is permissible. A monomer touching site j will have the weight $w_j = \exp(-\varepsilon_j/kT)$. This allows us to accommodate surface energetics, but it also allows the monomer to have an energy which is a function of position throughout the lattice. Our initial description will be for the case where each monomer has the same energy of attraction for a lattice site as any other, but later we will describe a simple generalization (see Eq. 7) which allows the energy to be monomer specific, that is to say $w_{j,k} = \exp(-\varepsilon_{j,k}/kT)$, where k is the ordinal location of the kth monomer in a linear chain of N monomers ($1 \leq k \leq N$) and j is the site number.

|    | 1  | 2  | 3  | 4  | 5  | 6  | 7  | 8  | 9  | 10  | 11  | 12  | 13  | 14  | 15  |
|----|----|----|----|----|----|----|----|----|----|-----|-----|-----|-----|-----|-----|
| 1  |    | $w_2$ |    |    |    | $w_6$ |    |    |    |     |     |     |     |     |     |
| 2  | $w_1$ |    | $w_3$ |    |    |    | $w_7$ |    |    |     |     |     |     |     |     |
| 3  |    | $w_2$ |    | $w_4$ |    |    |    | $w_8$ |    |     |     |     |     |     |     |
| 4  |    |    | $w_3$ |    | $w_5$ |    |    |    | $w_9$ |     |     |     |     |     |     |
| 5  |    |    |    | $w_4$ |    |    |    |    |    | $w_{10}$ |     |     |     |     |     |
| 6  | $w_1$ |    |    |    |    |    | $w_7$ |    |    |     | $w_{11}$ |     |     |     |     |
| 7  |    | $w_2$ |    |    |    | $w_6$ |    | $w_8$ |    |     |     | $w_{12}$ |     |     |     |
| 8  |    |    | $w_3$ |    |    |    | $w_7$ |    | $w_9$ |     |     |     | $w_{13}$ |     |     |
| 9  |    |    |    | $w_4$ |    |    |    | $w_8$ |    | $w_{10}$ |     |     |     | $w_{14}$ |     |
| 10 |    |    |    |    | $w_5$ |    |    |    | $w_9$ |     |     |     |     |     | $w_{15}$ |
| 11 |    |    |    |    |    | $w_6$ |    |    |    |     |     | $w_{12}$ |     |     |     |
| 12 |    |    |    |    |    |    | $w_7$ |    |    |     | $w_{11}$ |     | $w_{13}$ |     |     |
| 13 |    |    |    |    |    |    |    | $w_8$ |    |     |     | $w_{12}$ |     | $w_{14}$ |     |
| 14 |    |    |    |    |    |    |    |    | $w_9$ |     |     |     | $w_{13}$ |     | $w_{15}$ |
| 15 |    |    |    |    |    |    |    |    |    | $w_{10}$ |     |     |     | $w_{14}$ |     |

**Figure 2:** This figure describes the matrix used to solve the problem of a polymer molecule of N monomers confined to the lattice of figure 1. The linear dimension M of the matrix is equal to the number of lattice sites (M =15). The column labeled 7 has a weight $w_7$ at every location corresponding to a nearest neighbor site to site 7. Every other location in the column has a weight zero. Similarly, the column labeled j has a weight $w_j$ at every location corresponding to the nearest neighbor sites to site j, the other sites in the column each have the weight zero. The above statements are invariant to the way we label the sites. In the text we show that the partition function (sum over states) is obtained by multiplying the matrix by itself N-1 times along with the multiplication by proper fore and aft row and column vectors.

**Table 1: Polymer Interface Problems.**

1) kinetics, polymer absorption, absorption kinetics,
2) adsorption onto physically or chemically rough surfaces,
3) adsorption onto curved surfaces, into wedges, into fractal surfaces,
4) density variation at polymer interfaces,
5) surface tension, wetting,
6) adhesion and glue,
7) adsorption onto liquid-liquid, liquid-air and liquid-solid interfaces,
8) polymer rings, combs and stars onto or near a surface,
9) colloid stability, flocculation, water treatment,
10) polymer dispersants,
11) paper production,
12) rubber elasticity,
13) composite materials, polymer concrete, polymer sand,
14) nanocomposites, polymer-clay composites,
15) block-copolymer interfaces,
16) polymer blends, emulsification of immiscible polymer blends by adding block copolymers,
17) spinodal decomposition in polymer blends,
18) describing phase transitions that occur at a polymer interface, critical energy dependence on lattice type,
19) thin films, films as protective surfaces,
20) diffusion through and within membranes, from one polymer to another, improved battery operation,
21) semi-crystalline polymer morphology, Polymer crystal kinetics,
22) polymer dynamics at interfaces, enhanced mobility of confined polymers, polymer reptation at interfaces,
23) glass temperature of polymers confined to pores or in thin films as a function of pore size and polymer molecular weight,
24) lubrication and wear, viscoelasticity of confined polymers,
25) turbulence suppression near surfaces,
26) self-healing of fractured surfaces,
27) skin effect,
28) encapsulation, drug delivery,
29) chemical reactions on surface, surface enhanced catalysis,
30) reactivity and stability of surface, corrosion,
31) polymerization of thin films,
32) polymers covalently attached to surface, polymer brushes, grafted polymers,
33) tertiary oil recovery, enhanced oil recovery,
34) membrane sensors,
35) photo resists, polymer solar cells,
36) optical thin films,
37) metal-polymer interfaces (Cu-polyamide for example,
38) polymer surface modification with plasmas, plasma deposition of polymers,
39) decoration and surface finishing of plastics,
40) biopolymer threading a membrane transition,
41) biological membrane structure and phenomena, the golgi apparatus,
42) effect of confinement on the collapse of a protein to its globular form,
43) cell aggregation, tissue formation,
44) surface induced enzymatic activity,
45) gel permeation chromatography,
46) separation by flow,
47) thin layer chromatography,
48) neutron reflectivity,
49) field flow fractionation,
50) ellipsometry.